\documentclass{article}
\usepackage[utf8]{inputenc}
\usepackage{xcolor}
\usepackage{url}
\usepackage{authblk}
\usepackage{booktabs}
\usepackage{graphicx}
\usepackage{subcaption}
\usepackage{amsmath}
\usepackage{amsfonts}
\usepackage{amssymb}
\usepackage{hyperref}
\usepackage[capitalise]{cleveref}

\usepackage[ruled, linesnumbered, procnumbered]{algorithm2e}

\title{Automatic motion estimation with applications to hiPSC-CMs}
\author[1]{Henrik Finsberg}
\author[2]{Verena Charwat}
\author[3,4]{Kevin Healy}
\author[1]{Samuel T. Wall}
\affil[1]{Simula Research Laboratory,  Norway}
\affil[2]{Organos, Inc, Berkeley, CA, United States}
\affil[3]{Department of Bioengineering, University of California, Berkeley, CA, United States}
\affil[4]{Department of Materials Science and Engineering, College of Engineering, University of California, Berkeley, CA, United States}
\begin{document}
\maketitle

\begin{abstract}

Human induced pluripotent stem cell-derived cardiomyocytes (hiPSC-CMs) are an effective tool for studying cardiac function and disease, and hold promise for screening drug effects on human tissue. Changes to motion patterns in these cells are one of the important features to be characterized to understand how an introduced drug or disease may alter the human heart beat. However, quantifying motion accurately and efficiently from optical measurements using microscopy is currently lacking. In this work, we present a unified framework for performing motion analysis on a sequence of microscopically obtained images of tissues consisting of hiPSC-CMs. We provide validation of our developed software using a synthetic test case and show how it can be used to extract displacements and velocities in hiPSC-CM microtissues. Finally, we show how to apply the framework to quantify the effect of an inotropic compound. The described software system is distributed as a python package that is easy to install, well tested and can be integrated into any python workflow.
\end{abstract}

\section{Introduction}

Cardiovascular diseases account for 1 out of 3 deaths in the US \cite{virani2021heart}. At the same time, the yearly number of newly developed drugs is declining \cite{FORDYCE20151567}. 
One of the challenges in discovering new cardiac drugs has been the lack of in vitro screening systems.  However, the use of human induced pluripotent stem cell-derived cardiomyocytes (hiPSC-CMs) has opened a new path for drug development with great potential that can lead to faster and safer development of new drugs, reducing the number of animal experiments and allowing tailored drug development.\cite{ye2018application, ballan2020single}. 

Functional \emph{in vitro} assessment of cardiomyocytes is an important step in drug development, both for drug discovery and drug safety testing. Being able to screen at an early pre-clinical stage is critical, and in particular, robust methods for assessing mechanistic aspects of cardiac function are needed \cite{peters2012evaluation}.

In addition to studying the effect of novel drugs, hiPSC-CMs can be used to study genetic cardiac diseases, several of which can alter not only electrophysiology but also mechanical contractile behavior \cite{kiviaho2015distinct, birket2015contractile}. It is therefore important to be able to quantify the cardiac motion in a robust and unbiased fashion and with reliable precision. 
Several methods already exist, some of which are based on \emph{in vitro} force measurements. For example, atomic force microscopy (AFM) is a method for determining mechanical properties with high resolution by placing a mechanical probe on the tissue and measuring the movement and resulting force generated by the tissue. This method has been used to examine the inotropic and chronotropic responses of cardiomyocyte contraction induced by various pharmacological interventions in \cite{chang2013characterization}. However, since AFM is applied at a single point, several experiments have to be conducted to obtain spatial resolution. Furthermore, elasticity and height at the contact point may change during a beat, which might lead to heterogeneity in the measured forces \cite{liu2012atomic}.

Other types of non-invasive methods are based on video-microscopy, where the cells are recorded under a microscope and cellular displacements are computed by applying techniques from computer vision \cite{hayakawa2014image,czirok2017optical, huebsch2015automated}. One such technique, known as block matching was successfully applied in \cite{huebsch2015automated} to estimate motion vectors of macroblocks in brightfield and fluorescence videos of hiPSC-CM derived microtissues. Here the authors also show how motion tracking software can be used to assess baseline behavior and drug response of hiPCS-CMs. In \cite{ribeiro2017multi} the preparations contained fluorescent microbeads on micropatterend hIPCS-CM monolayers, and contractile parameters were successfully measured.

Digital image correlation (DIC) is another technique that can be used on video-microscopy data. In DIC, patterns that are referred to as speckles are tracked and surface displacements are computed based on the motion of these speckles. In \cite{ahola2014video}, DIC was used to extract beating characteristics from single cells of induced pluripotent stem cell-derived cardiomyocytes that had no clear movement axis. Machine learning methods are also emerging as potential methods to automatically detect and predict the mechanistic action of unknown cardioactive drugs.  \cite{lee2015machine, lee2017machine}

A common problem in motion analysis of hIPCS-CM is the availablity of developed methods to other users and groups.  The software used to perform the motion analysis is often either not public, or it is implemented in a proprietary language (such as Matlab) which can make it largely inaccessible for many users. 

Recently, there has been move towards open source alternatives in video-microscopy \cite{sala2018musclemotion, scalzo2021dense}. Here we propose a pure python library that can easily be installed and which can work with regular python constructs such as NumPy arrays. This makes it  simple to integrate into an existing data analysis pipeline. The library uses well-established methods from computer vision for motion tracking and enables multi-core parallelism and out-of-core computations for high-scale and high-throughput data analysis. We also provide a simple graphical user interface that can be used for analyzing a single dataset.

In this paper, we will demonstrate the different capabilities of this motion tracking software, and discuss different types of challenges and potential solutions when applying motion analysis to datasets of hIPCS-CM.

\section{Methods}
\label{sec:methods}

\subsection{Data sources}
\label{sec:data_sources}
Each video file consists of a sequence of images $I(x_m, y_n, t_k)$ for $ 1 \leq m \leq M, 1 \leq n \leq N, 1 \leq k \leq T$ with $M$ and $N$ being respectively the width and height in pixels, and $T$ being the number of frames in the stack. In addition, the images contain metadata with the time stamps and a conversion factor to convert from pixel units to physical units (i.e $\mu$m).

\subsection{Motion estimation}
\label{sec:motion_estimation}

\subsubsection{Governing equations}
A video recording of a collection of cells can formally be described as a sequence of images. Let $I(x, y, t)$ denote the image sequence at position $(x, y)$ and time $t$. At a later time, $t + \Delta t$, the cells may have moved to a new position and hence a pixel at $(x, y)$ has now moved to $(x + \Delta x, y + \Delta y)$, i.e
\begin{align}
    I(x, y, t) = I(x + \Delta x, y + \Delta y, t + \Delta t).
\end{align}
By applying a Taylor expansion and dropping higher order terms, we end up with 
\begin{align}
    I(x, y, t) 
    &= I(x + \Delta x, y + \Delta y, t + \Delta t) \\
    &\approx I(x, y, t) + \frac{\partial I}{\partial x} \Delta x + \frac{\partial I}{\partial y} \Delta y + \frac{\partial I}{\partial t} \Delta t
\end{align}
\begin{align}
    \implies  \frac{\partial I}{\partial x} \Delta x + \frac{\partial I}{\partial y} \Delta y + \frac{\partial I}{\partial t} \Delta t = 0
\end{align}
and in the limit, we get the optical flow equation
\begin{align}
    \frac{\partial I}{\partial x} \cdot \frac{\partial x}{\partial t} +  \frac{\partial I}{\partial y} \cdot \frac{\partial y}{\partial t}= - \frac{\partial I}{\partial t}.
    \label{eq:optical_flow}
\end{align}
Here $\frac{\partial I}{\partial x}$, $\frac{\partial I}{\partial y}$ and $\frac{\partial I}{\partial t}$ are respectively the image gradients along the spatial dimensions $x$ and $y$ and the temporal dimension $t$, while $\frac{\partial x}{\partial t}$ and $\frac{\partial y}{\partial t}$ are unknowns that can be interpreted as the motion in the $x-$ and $y$ direction respectively relative to the chosen reference image.

In a discrete setting we pick the image a time $t_{\text{ref}}$ as reference, i.e $I(x, y, t_{\text{ref}})$. Let us denote the motion relative to this image by $\mathbf{u}(x, y, t; t_{\text{ref}})$. Then $\mathbf{u}$ is a vector field whose components are the unknowns in the optical flow equation, i.e
\begin{align}
    \mathbf{u}(x, y, t; {\text{ref}}) &= 
    \begin{pmatrix}
    u_x(x, y, t; t_{\text{ref}}) \\ u_y(x, y, t; t_{\text{ref}})
    \end{pmatrix} \\ &= 
    \begin{pmatrix}
    \frac{\partial x}{\partial t}(x, y, t; t_{\text{ref}}) \\ \frac{\partial y}{\partial t}(x, y, t; t_{\text{ref}}),
    \end{pmatrix}
\end{align}
where $\frac{\partial x}{\partial t}$ and $\frac{\partial y}{\partial t}$ are the solution of the optical flow equation from time $t_{\text{ref}}$ to $t$. For a fixed reference image, we will refer to $\mathbf{u}$ as the \emph{displacement field}. Note that the motion from one frame to itself is zero, i.e
\begin{align}
    \mathbf{u}(x, y, t_{\text{ref}}; t_{\text{ref}}) = \mathbf{0}.
\end{align}
Instead of using a fixed reference image, the reference image can be a function of the current image. For example, we can pick the image that is $s$ number of frames before the current frame and compute the motion from from $I(x, y, t_{k-s})$ to $I(x, y, t_{k})$. Dividing by the time between these two frames then gives an approximation of the \emph{velocity field},
\begin{align}
\mathbf{v}(x, y, t_k; s) = \frac{\mathbf{u}(x, y, t_k; t_{k-s})}{t_k - t_{k-s}}.
\label{eq:velocity_motion}
\end{align}
We will refer to $s$ as the \emph{spacing}. The velocity field can also be computed from a known the displacement field using
\begin{align*}
    \mathbf{v}(x, y, t_k; s, t_{\text{ref}}) &= \frac{\mathbf{u}(x, y, t_k; t_{\text{ref}}) - \mathbf{u}(x, y, t_{k-s}; t_{\text{ref}})}{t_k - t_{k-s}}.
\end{align*}
Note however, that this result will depend on the choice of reference frame used in the displacement field, since the solution of the optical flow computations are in general not transitive \cite{christensen2001consistent}, i.e $\mathbf{u}(x, y, t_k; t_s) + \mathbf{u}(x, y, t_s; t_k) \neq  \mathbf{0}$ for $k \neq s$.

\subsubsection{Dimentionality reduction}
The fields $\mathbf{u}$ and $\mathbf{v}$ contain a vector for each pixel in the sequence of frames. We will refer to this type of object as \emph{VectorFrameSequence}. We can reduce this to a \emph{FrameSequence}, i.e a sequence of frames with a scalar value for each pixel, in several ways. For example, one can take the $x$- or $y$ component, or we can compute the norm of the vectors at each pixel, i.e

\begin{align}
    u_{\|\cdot \|_{\ell^2}}(x, y, t; t_{\text{ref}}) = \sqrt{u_x(x, y, t; t_{\text{ref}})^2 + u_x(x, y, t; t_{\text{ref}})^2}.
\end{align}

A FrameSequence can be further reduced along the spatial dimension by averaging over the pixels, i.e
\begin{align}
    \bar{u}_{\|\cdot \|_{\ell^2}}(t, ; t_{\text{ref}}) = \frac{1}{MN}\sum_{m=1}^M\sum_{n=1}^N  u_{\|\cdot \|_{\ell^2}}(x_m, y_n, t; t_{\text{ref}}).
\label{eq:trace_spatial_red}
\end{align}
Similarly, reduction along the temporal dimension can be done by for example computing the maximum value for a given pixel
\begin{align}
    u_{\|\cdot \|_{\ell^2}}^{\max}(x, y; t_{\text{ref}}) = \max_{0 \leq t \leq T} u_{\|\cdot \|_{\ell^2}}(x, y, t; t_{\text{ref}}).
\end{align}
Finally, a pixel mask can also be included, to e.g filter out pixels that do not contain cells and which should not be included when computing the average displacement. One way to construct such a mask is to select those pixels where the maximum attained displacement over all time steps is smaller than some threshold value $\epsilon$, e.g
\begin{align}
    \mathcal{M} = \left\{ (x, y) : u_{\|\cdot \|_{\ell^2}}^{\max}(x, y) <  \epsilon \right\},
\label{eq:mask}
\end{align}
in which case e.g \eqref{eq:trace_spatial_red} becomes
\begin{align}
    \bar{u}_{\|\cdot \|_{\ell^2}}(t; ; t_{\text{ref}}) = \frac{1}{|\mathcal{M}|}\sum_{(x, y) in \mathcal{M}} u_{\|\cdot \|_{\ell^2}}(x_m, y_n, t; t_{\text{ref}}).
\label{eq:trace_spatial_red_mask}
\end{align}
The value of $\epsilon$ needs to be chosen ad-hoc, which can be problematic. We found that choosing
\begin{align}
    \epsilon = \frac{1}{MN} \sum_{m=1}^M\sum_{n=1}^N u_{\|\cdot \|_{\ell^2}}^{\max}(x_m, y_n),
\label{eq:mask_epsilon}
\end{align}
gave a reasonable mask and this will be used by default in our results.

\subsubsection{Solving the optical flow equation}
The optical flow problem has two unknowns; one component in the $x$-direction and one component in the $y$-direction. Therefore one more equation is needed to find a unique solution. 

The incorporation of the second equation needed to solve the optical flow problem is handled differently in different algorithms.
We will first examine a few standard algorithms for motion analysis implemented in the open-source computer vision software OpenCV\cite{bradski2000opencv}. In particular the Lucas-Kanade method\cite{lucas1981iterative}, the Farnebäck method\cite{farneback2003two}, and Dual TV-$L^1$ \cite{zach2007duality}. In addition, we also test an in-house implementation of a block matching algorithm, similar to the one used in \cite{huebsch2015automated}. Later we will select the Farnebäck method as the default algorithm for motion analysis and conduct the remaining experiments using this algorithm.

All of these algorithms represent different types of motion estimators. The Lucas-Kanade and the block matching algorithm are sparse methods that only estimate the motion in a subset of the data, and hence the output of these methods have a lower resolution than the input data. As a result, we also perform an interpolation step on the output, so that the resulting motion vectors have the same dimensions regardless of the methods used. The Farnebäck and Dual TV-$L^1$ on the other hand are dense methods that will estimate the motion at each pixel and therefore no interpolation is needed.

\subsubsection{Software and implementation details}
\label{sec:soft_impl_details}
All the code for performing motion analysis is packaged in a library called \emph{mps-motion} and is publicly available\footnote{Source code is available at \url{https://github.com/ComputationalPhysiology/automatic-motion-estimation-hiPSC-CM}} . The package is implemented in python and uses OpenCV\cite{bradski2000opencv} for most of the computer-vision related tasks.

Algorithm \ref{algo:motion} lists the full algorithm that takes a stack of images together with the time stamps, runs the motion analysis and outputs traces of the average displacement and velocity norm. 

\begin{algorithm}
    \caption{Motion analysis algorithm}\label{algo:motion}
    \KwIn{$I=I(x_m, y_n, t_k), 1 \leq m \leq M, 1 \leq n \leq N, 1 \leq k \leq T$ -- Image sequence }
    \KwIn{$t$ -- time stamps}
    \KwIn{method -- Algorithm for finding optical flow, by default Farnebäck}
    \KwIn{scale -- Scale factor for down-sampling images}
    \KwIn{spacing -- Image spacing used in velocity computation}
    \KwOut{$u$ -- Average displacement norm}
    \KwOut{$v$ -- Average velocity norm}

    $I = \texttt{resize}(I, \text{scale})$  -- Down-sample images\\
    $\mathbf{v} = \texttt{optical\_flow}\_v(I, \text{method}, \text{spacing})$ -- Compute velocity ($\mathbf{v}$)\\
    $v$\_tmp = $\texttt{compute\_mean\_norm}(\mathbf{v})$\\
    ref\_idx = $\texttt{estimateRefereceImage($t, v$\_tmp)}$ \\
    $\mathbf{u} = \texttt{optical\_flow}\_u(I, \text{method}, \text{ref\_idx})$ -- Compute displacement ($\mathbf{u}$)\\
    $\mathcal{M} = \texttt{find\_mask}(\mathbf{u})$ -- Compute mask used for filtering \\
    $\mathbf{u} = \texttt{apply\_mask}(\mathbf{u}, \mathcal{M})$ -- Apply filtering\\
    $\mathbf{v} = \texttt{apply\_mask}(\mathbf{v}, \mathcal{M})$  -- Apply filtering\\
    $v$ = $\texttt{compute\_mean\_norm}(\mathbf{v})$ -- Average velocity norm\\
    $u$ = $\texttt{compute\_mean\_norm}(\mathbf{u})$ -- Average displacement norm\\
\end{algorithm}

First, a down-sampling of the images is performed with a chosen scaling factor. In \cref{sec:downsampling} we investigate the accuracy and speed when using different scaling factors. Next step is to estimate the reference index using Algorithm \ref{fun:estimateRefereceImage} which is further explained in \cref{sec:reference_frame}, and requires that we first estimate the velocity fields. After specifying the reference index we again run the optical flow algorithm to estimate the displacement. Once we have the displacement field we compute the mask using \cref{eq:mask} and apply this to the displacement. We also apply the same mask to the already computed velocity field. Finally we compute the average norm of the displacement and velocity.

\subsubsection{Parallel computations}
Computing the optical flow and the mechanical features are both CPU-intensive and memory-demanding operations. For example, the displacement field for an image sequence with 267 images and each image being $2044 \times 1174$ pixels, has a of shape $(2044, 1174, 267, 2)$. These displacement values with double precision would take up more than 10GB of memory just to store the values in memory. Performing computations on such large arrays will therefore push the memory limit of most modern laptops. To overcome this limitation we use a library called Dask\cite{rocklin2015dask} which enables parallel and out-of-core computations.

\section{Results}

\subsection{Verification}
\label{sec:verification}
To verify that the implementation is correct, we compare the resulting displacement obtained from the motion tracking software with a known solution. To generate a known solution, we take a single frame and generate another frame by warping it according to some known function, $f: \mathbb{R}^2 \mapsto \mathbb{R}^2$. 
We used the following function to define the synthetic image motion

\begin{equation}
    f = f(x, y; a, b) = \begin{pmatrix}
        a x \\
        b y
    \end{pmatrix}.
    \label{eq:verification}
\end{equation}

with the following three parameter sets
\begin{align}
(a, b) \in \{ (0.005, 0.0), (0.0, 0.007), (0.005, 0.007) \},
\end{align} 
and we will refer to these synthetic data sets as $x, y$ and $xy$ respectively as they represent synthetic displacements along those respective axes. 

\begin{figure}
    \includegraphics[width=\textwidth]{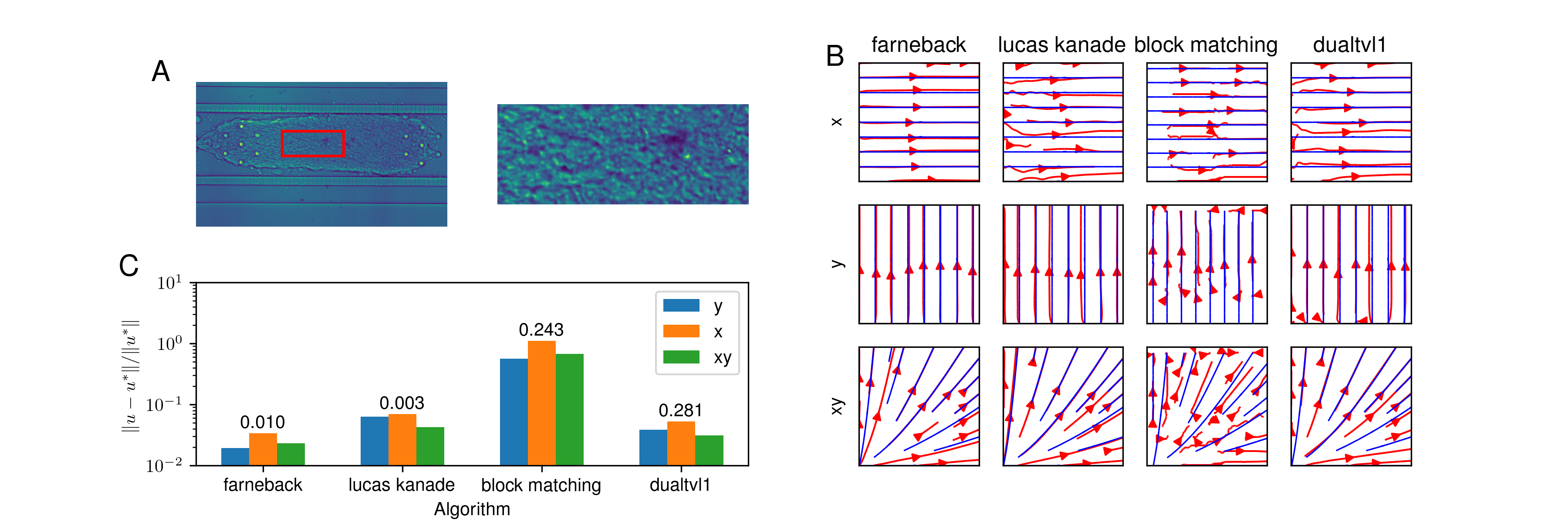}
    \caption{\label{fig:verification}\textbf{A}: Selected region of microscopic image used for generating synthetic motion data. The left panel shows the entire image of the microtissue with the selected region highlighted as a red box. The right panel shows a zoom-in on the selected region. \textbf{B}: Estimated motion compared to exact solutions. The estimated displacement fields are shown as red streamlines while the exact solutions are shown as blue streamlines. Each row represent a different data set and each column is the results obtained from a specific algorithm.\textbf{C}: Relative $L^2$ error between estimated and true displacement using the different optical flow algorithms for different synthetic displacement. The numbers above the orange bars are the average number of seconds it took for running the algorithm.}
\end{figure}

The image used for the verification is shown in panel  A in \cref{fig:verification}. The image that is warped is a selected region taken from a microscopic image of an experimental hiPSC-CM system.

Results are shown in Panel B in \cref{fig:verification}.  The resulting optical flow patterns are shown as red streamlines for each the four described optical flow algorithms. The exact solutions are shown as blue streamlines. We see that all algorithms capture the essential motion patterns that were synthesized, but the accuracy varies to some degree. In panel C the relative $L^2$-error between the estimated and the exact optical flow is shown. We notice that the error in the block matching algorithm is much higher than in the other algorithms, while the other algorithms have more or less the same $L^2$ error.  The Farnebäck method gave the overall smallest error in all cases, and was the second fastest of the tested algorithms. For the remaining of this study we will only use the Farnebäck method method.

\subsection{Displacement and velocity estimation}
In this section, we use the motion data generated from two experimental recordings and investigate the sensitivity to different options that can be set during motion estimation.

The two image stacks come from two different hiPSC-CM microtissue studies, and were acquired with different optical instruments. In that sense, this assessment is intended capture variation seen by using different experimental setups.
Some relevant metadata for these chips are listed in \cref{tab:chip_metadata}.

\begin{table}
\centering
\begin{tabular}{lrr}
\toprule
Name & Dataset 1 & Dataset 2\\
\midrule
size & 2044 $\times$ 1174 pixels&1024 $\times$ 205 pixels \\
number of frames & 267&1500 \\
frames per seconds & 84&200 \\
duration & 3172 ms&7511 ms \\
physical unit & 0.325 $\mu$m / pixel&1.3552 $\mu$m / pixel \\
\bottomrule
\end{tabular}
\caption{Metadata for the chips used in the analysis}
\label{tab:chip_metadata}
\end{table}

\begin{figure}
    \includegraphics[width=\textwidth]{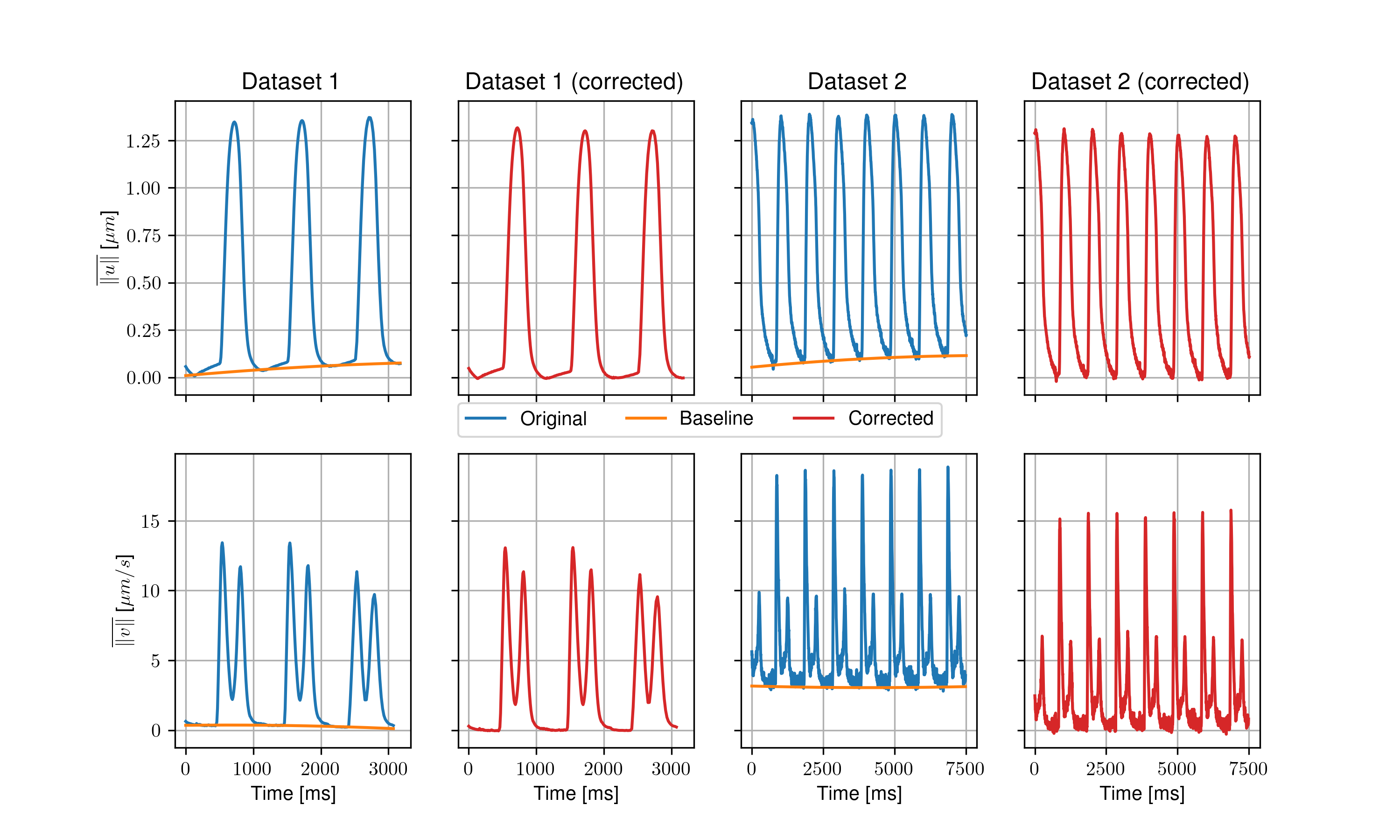}
    \caption{\label{fig:mean_traces} Average displacement norm (top row) and velocity norm (bottom row) computed using \eqref{eq:trace_spatial_red_mask}, i.e averaged over the masked area containing cells where the maximum displacement were larger than the average maximum displacement. First and third columns shows the original average traces for Dataset 1 and 2 together with the estimated baseline. Second and forth column shows the same datasets, but with the baseline subtracted from the average traces.}
\end{figure}

For both datasets, we estimated the reference frame using Algorithm \ref{fun:estimateRefereceImage}, using a spacing of 5 frames in the velocity computations. More details around the estimation of the reference frames are given in \cref{sec:reference_frame}. We also applied a filter where all pixels with a maximum displacement less than the mean maximum displacement were set to zero, i.e using \eqref{eq:mask_epsilon}. We noticed a drifting in the baseline values for the displacement, which translate into a non-zero baseline for the velocity traces. We can remove this baseline by applying a baseline estimation algorithm \cite{mazet2005background} and subtracting this baseline from the signal, see \cref{fig:mean_traces}. The resulting average displacement and velocity norm for the two datasets are shown in \cref{fig:mean_traces}. Here we also show the traces with and without the baseline correction.

\begin{figure}
    \includegraphics[width=\textwidth]{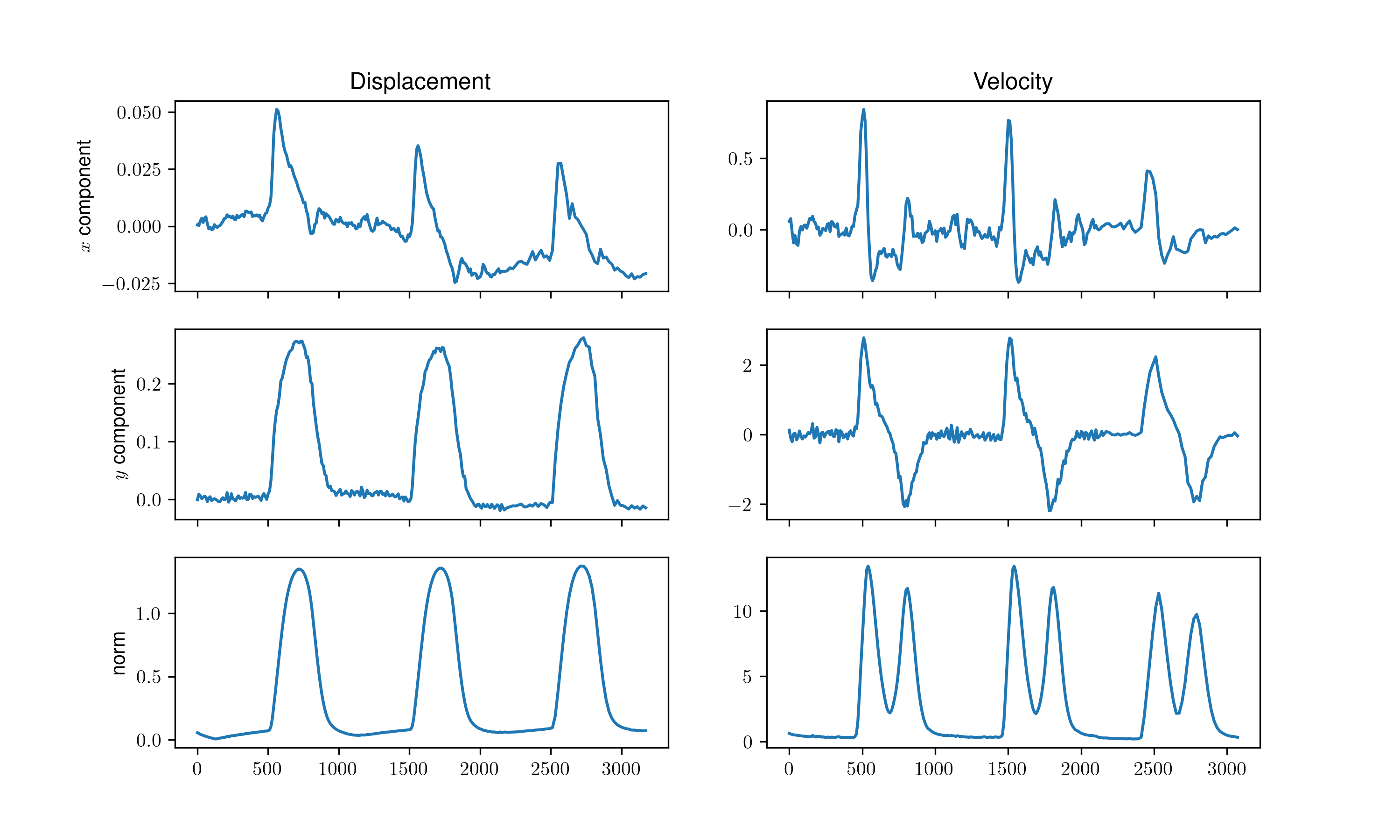}
    \caption{\label{fig:mean_comp_traces} Average displacement (left column) and velocity (right column) for Dataset 1. First, second and third rows shows respectively the $x$-component, $y-$component and the norm.}
\end{figure}

Note that it is also possible to extract the $x-$ and $y-$ contribution to the displacement and velocity as shown in \cref{fig:mean_comp_traces} for Dataset 1.

From these traces, there are three distinct beats in Dataset 1 and 7 beats in Dataset 2. We  notice that the traces obtained from Dataset 2 contains more noise, especially in the components of the velocity.  For the velocity, we see the characteristic two peaks for each beat representing the maximum contraction speed and maximum relaxation speed respectively. 

\begin{figure}
    \includegraphics[width=\textwidth]{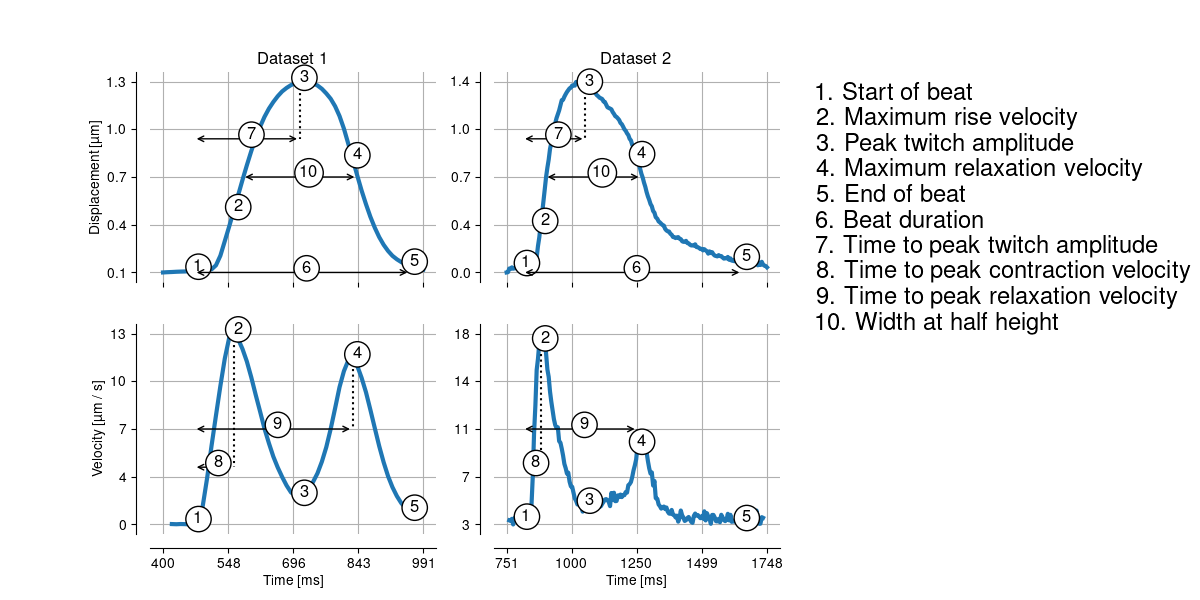}
    \caption{Features used to characterize the displacement and velocity of muscle twitches. The top and bottom row show respectively the characteristics of the displacement and velocity with traces taken from the first beat of Dataset 1 (left) and 2 (right). \label{fig:features}}
\end{figure}

In \cref{fig:features} we have extracted the first beat for the displacement and the velocity for both data sets and attached a few commonly used features that can be used to quantify the motion. \cref{tab:motion_features} list the different values of these features for this particular beat for the two data sets.

\begin{table}
\centering
\begin{tabular}{lrr}
\toprule
Name & Dataset 1 & Dataset 2 \\
\midrule
Maximum rise velocity & 12.0 & 14.7 \\
Peak twitch amplitude & 1.0 & 1.0 \\
Maximum relaxation velocity & 9.6 & 5.6 \\
Beat duration & 450.3 & 541.0 \\
Time to peak twitch amplitude & 0.0 & 266.0 \\
Time to peak contraction velocity & 140.1 & 125.0 \\
Time to peak relaxation velocity & 410.3 & 506.0 \\
Width at half height & 215.6 & 329.9 \\
\bottomrule
\end{tabular}
\caption{Values for different features for Dataset 1 and 2 shown in \cref{fig:features}}
\label{tab:motion_features}
\end{table}

\subsubsection{Effect of filtering pixels \label{sec:filtering}}
\begin{figure}
    \includegraphics[width=\textwidth]{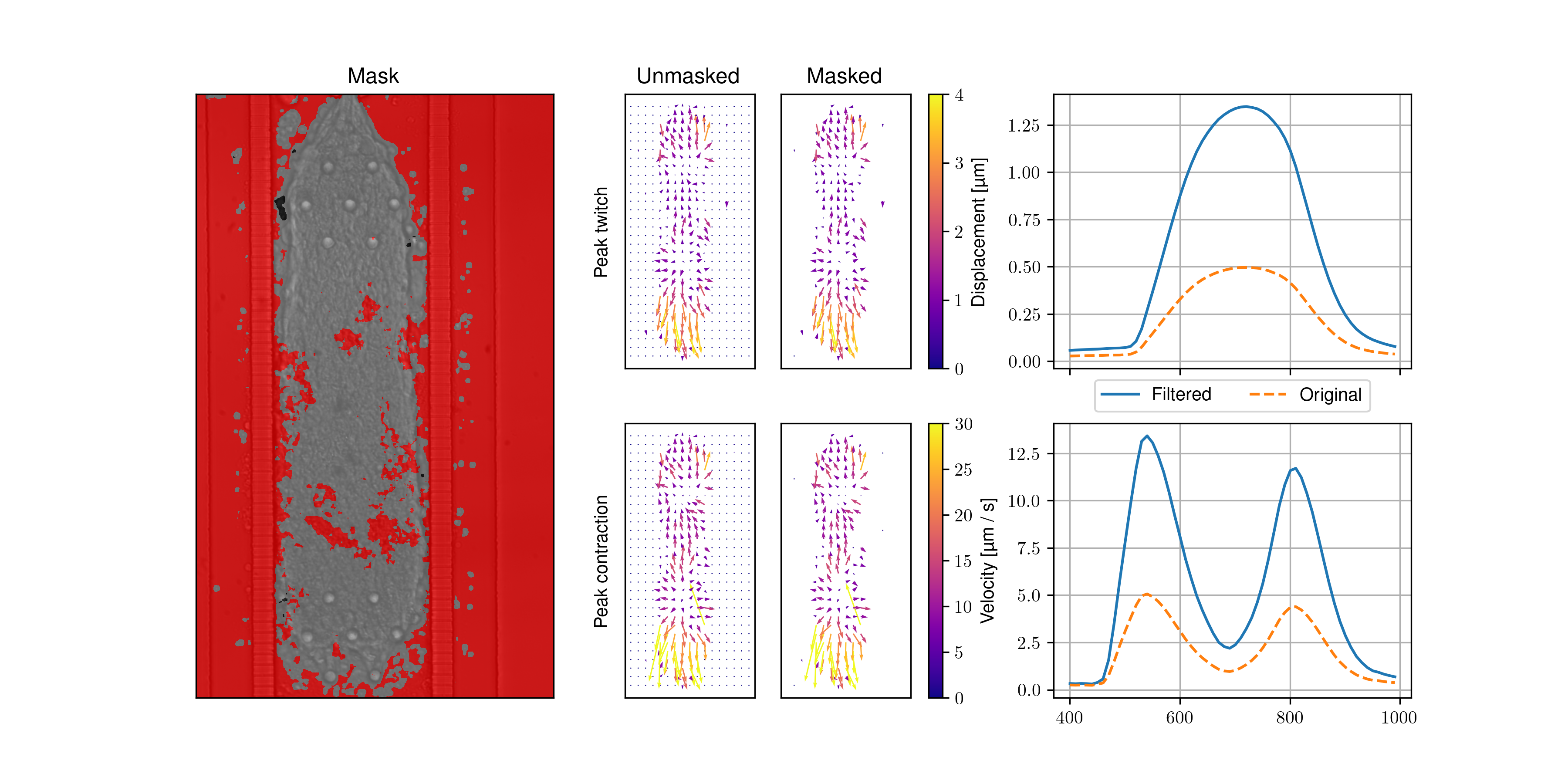}
    \caption{Effect of filtering for Dataset 1. The first column shows the mask, where pixels in the red region have their displacements set to zero. Column 2 and 3 shows the unmasked and masked vectors respectively for the displacement vectors at the time of peak twitch at the top and the velocity vectors at the time of peak contraction at the bottom. Right most column show the resulting average traces of the norm of the vectors using the original and filtered vectors \label{fig:masked_filter}}
\end{figure}

As outlined in \cref{sec:soft_impl_details} we applied a filtering to the displacement and velocity to remove pixels with a low maximum displacement. In \cref{fig:masked_filter} we show the resulting displacement and velocity vectors, with and without filtering for Dataset 1. Here the filtering mask, defined in \eqref{eq:mask} and \eqref{eq:mask_epsilon}, contain pixels where the maximum displacement norm is smaller than the mean the maximum displacement norm, averaged over all the pixels.

It is evident that the displacement vectors are not heavily affected by this filtering, while for the velocity vectors we see that a large portion of the vectors outside of the tissue is removed during the filtering procedure.

\cref{fig:masked_filter} also show the resulting mean traces obtained with and without filtering applied. When applying the filtering, pixels with low displacement magnitude are excluded from the mean computation. The means are now computed over the pixels with higher displacement values and it will thus result in a higher mean displacement.

\subsubsection{Effect of down-sampling}
\label{sec:downsampling}
The computational effort to analyze the images depends on the resolution of the images and the number of time steps. Brightfield images used in motion analysis are typically high-resolution images. One option to reduce analysis time is to first perform a down-sampling of the images. However, during the down-sampling, the resulting images will be more blurry, and therefore it is important to understand the trade-off between speedup and loss of information about the motion. We define the relative error for a given signal as the error between the current signal and the signal at full scale, i.e

\begin{align}
    \mathrm{err}(w) = \frac{\| \bar{w}_{\|\cdot \|_{\ell^2}} - \bar{w}_{\|\cdot \|_{\ell^2}}^{100\%} \|_{\ell^2}}{\| \bar{w}_{\|\cdot \|_{\ell^2}}^{100\%} \|_{\ell^2}}
\end{align}

where $\bar{w}_{\|\cdot \|_{\ell^2}}$ is given by \eqref{eq:trace_spatial_red_mask} and superscript $100\%$ refers to the signal at full scale.
In \cref{tab:downsampling_error} we show the this error for the displacement $u$ and velocity $v$ at the different down-sampling rates.

\begin{table}
    \centering
    \begin{tabular}{l|ccccccccc}
        \toprule
        Scale & 0.10 & 0.20 & 0.30 & 0.40 & 0.50 & 0.60 & 0.70 & 0.80 & 0.90 \\
        \midrule
        $\mathrm{err}(u)$ (Dataset 1) & 0.10 & 0.03 & 0.02 & 0.02 & 0.02 & 0.02 & 0.02 & 0.02 & 0.01\\
        $\mathrm{err}(v)$ (Dataset 1)& 0.27 & 0.06 & 0.04 & 0.02 & 0.01 & 0.01 & 0.01 & 0.01 & 0.01 \\
        $\mathrm{err}(u)$ (Dataset 2) & 0.96 & 0.43 & 0.26 & 0.06 & 0.09 & 0.02 & 0.03 & 0.02 & 0.02\\
        $\mathrm{err}(v)$ (Dataset 2)& 3.08 & 1.58 & 0.90 & 0.46 & 0.08 & 0.09 & 0.04 & 0.03 & 0.02\\
        \bottomrule
    \end{tabular}
    \caption{Relative error compared to the trace at full scale in different levels of down-sampling. The top two rows represent results using Dataset 1, while the bottom two rows represent results using Dataset 2.}
    \label{tab:downsampling_error}
\end{table}

We see that there are some differences between the two datasets. As we can see from \cref{tab:chip_metadata}, Dataset 1 has more pixels, and we can consequently also downsample this dataset more without losing much accuracy. For both datasets, the displacement traces are within a 10\% relative error at scales down to 40\%. However, for the velocity traces, we see that the relative error is below 10\% for both data sets at scales higher than 50\%.

\begin{figure}
    \includegraphics[width=\textwidth]{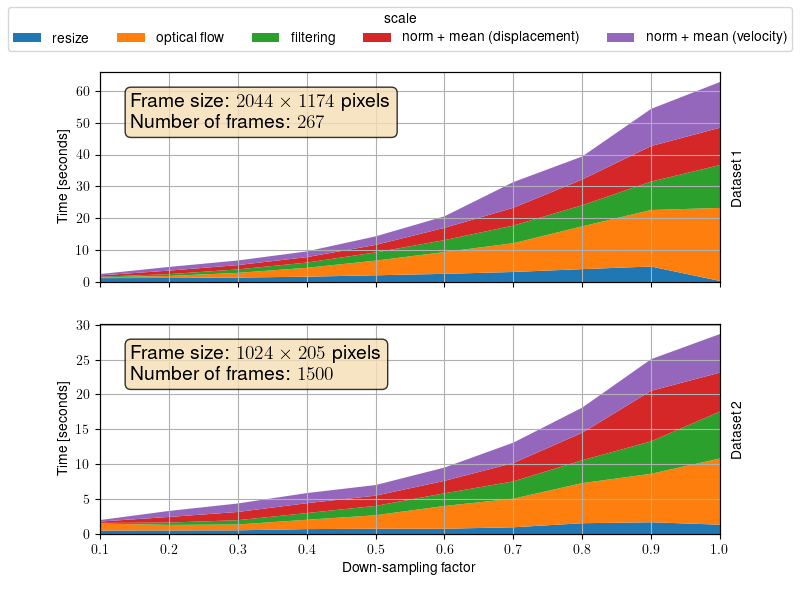}
    \caption{Timings in seconds for the different operations during motion analysis. \emph{Resize} is the time spent down-sampling the data from the original size to a lower resolution. \emph{Optical flow} is the time spent running the Farnebäck method, \emph{Filtering} is the time spent computing the mask and applying the filtering to both displacement and velocity. Finally, \emph{norm + mean (X)} is the time spent computing the norm and mean of all the arrays for \emph{X} (\emph{X} being velocity and displacement). \label{fig:timings_scale}}. These computations were performed on a MacBook Pro M1 Max with 64GB of memory.
\end{figure}

Down-sampling can also significantly decrease computational time, and \cref{fig:timings_scale} shows a stack plot of the timings for the different operations required in analysis using different levels of down-sampling. These operations include a resize of the stack of images, computation of the optical flow, and then postprocessing to compute and apply a mask, and finally computation of the norm and the mean trace of the displacement and the velocity. 

The total time for a full analysis is the sum of all the stacks which was 56 seconds and 28 seconds respectively for Datasets 1 and 2 using full resolution. At 50\% resolution the total time for a full analysis were 13 and 7 seconds respectively for Datasets 1 and 2. Out of this, about 30\% of the time is spent computing the optical flow, while the remaining time is divided evenly between computing the mean displacement and velocity, and performing the filtering. Note that the time it takes for the resizing at scale 1.0 is about 1 seconds. This is because a copy of all the frames are made when running the resize function in any case.

\subsection{Measuring drug effects \label{sec:drug_effect}}
In this section we apply the motion analysis to a dose escalation study using two different drugs; Omecamtiv mecarbil, a cardiac myosin activator, and Bay K8644, a selective L-Type Ca2+ channel agonist. Omecamtiv mecarbil was tested for doses of 1nM, 10nM, 100nM and 1000nM while Bay K8644was tested for doses of 10nM, 100nM and 1000nM.

\begin{figure}
   \includegraphics[width=\textwidth]{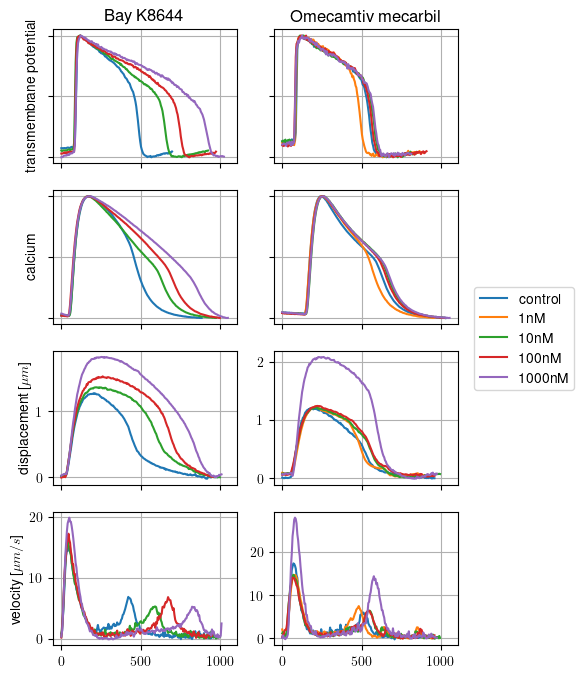}
    \caption{\label{fig:full_traces_drug} One beat of the transmembrane potential (first row), calcium transient (second row), displacement (third row) and velocity (fourth row) for a drug escalation study using Bay K8644 (left column) and Omecamtiv mecarbil (right column ). The different traces within each panel represents different doses of the drug. For transmembrane potential and calcium, the traces are normalized. }
\end{figure}

In \cref{fig:full_traces_drug} we show the resulting transmembrane potential, intracellular calcium transient, displacement, and velocity traces for the two drugs. Transmembrane potential and calcium trace are extracted from averaging the pixel values in the fluorescence images, and one representative beat is extracted and aligned at the upstroke. The displacement and velocity are estimated from brightfield images using Algorithm \ref{algo:motion}.

\begin{figure}
   \includegraphics[width=\textwidth]{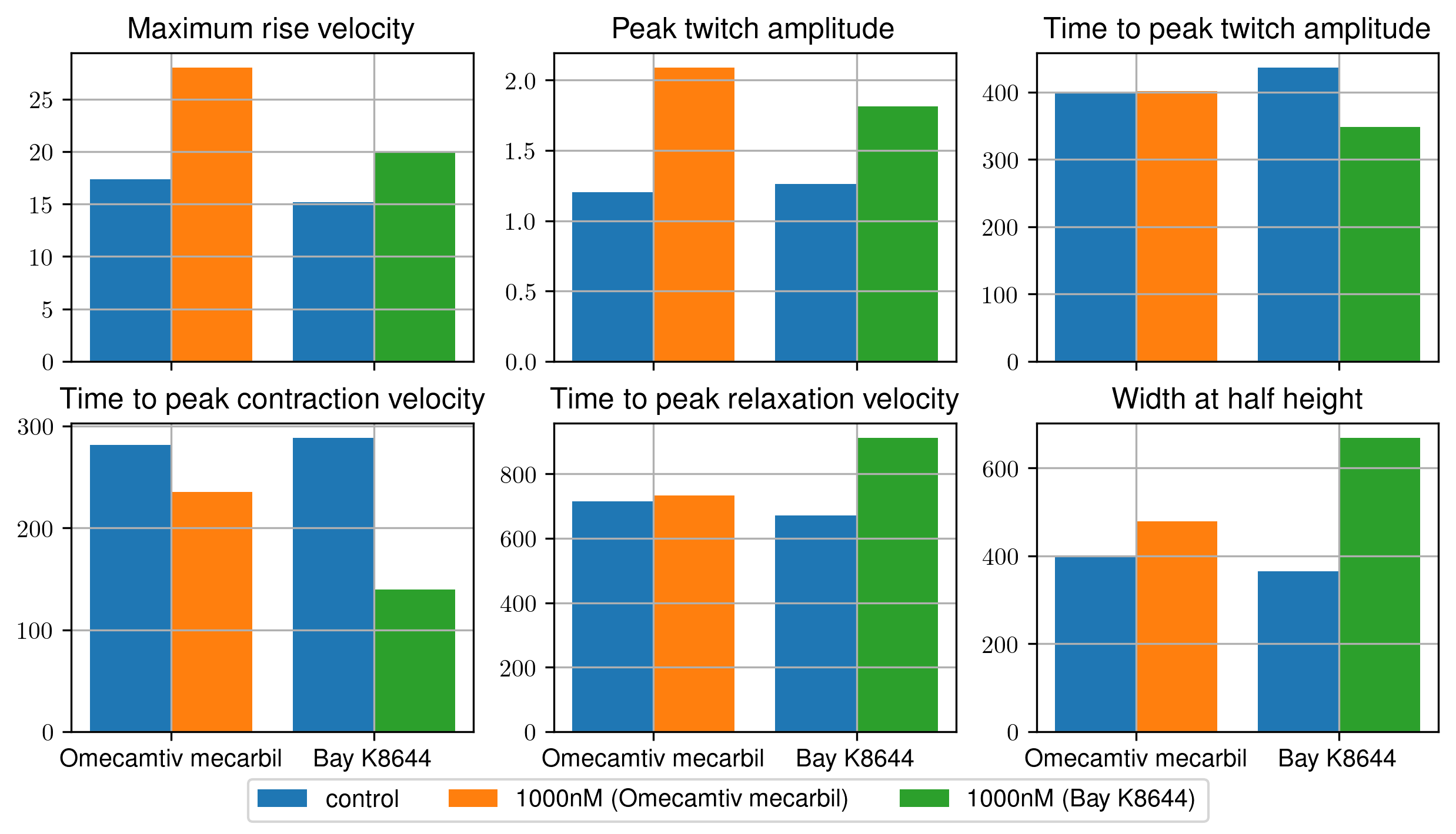}
    \caption{\label{fig:drug_features}Features described in \cref{fig:features} for Omecamtiv mecarbil and Bay K8644 using the control and the highest dose. In both cases the highest doses were 1000nM.}
\end{figure}

In \cref{fig:drug_features} we plot the features described in \cref{fig:features} for the two drugs, by comparing the control with the highest dose.
It is clear that both drugs increase the peak twitch amplitude as well as the width of the displacement traces at half height. We also see a decline in the time-to-peak contraction velocity, and time-to-peak twitch amplitude for an increased dose of Bay K8644, while this seems to remain more or less constant for Omecamtiv mecarbil.

\section{Discussion}
We have designed an efficient pipeline for conducting motion analysis of a sequence of tissue images, obtained microscopically from beating hiPSC-CMs. To ensure accuracy, we validated the software using a synthetic dataset and tested it against four different optical flow algorithms, as described in \cref{sec:verification}. Our analysis showed that the Farnebäck method was the most effective among the four algorithms, providing the best agreement with the validation data and the second best in terms of computing time.

Next, we assessed various components of the motion analysis pipeline using two distinct datasets generated from different experimental setups. One critical area of focus was applying a filter mask to eliminate pixels that do not contain cells contributing to the displacement (\cref{sec:filtering}). This is particularly important when the image contains numerous non-cellular pixels. However, it should be noted that this filter may also exclude cells that are potentially dead, which may not be the intended outcome. As a result, applying a mask may cause an overall increase in displacement and velocity."

To estimate the displacement field, optical flow computations use a fixed reference frame, while the reference frame for the velocity field depends on the frame of interest. In \cref{sec:spacing}, we conducted a sensitivity analysis on the delay between the current frame and the reference frame, which we termed as spacing in velocity calculations. If the spacing is too low, the velocity signal will be noisy, while a spacing that is too high will result in the peak velocity being averaged out, causing information loss. Our findings indicate that choosing a spacing corresponding to a physical time between 25 and 100 milliseconds offers a good balance between accuracy and noise accumulation.

We developed an automatic method to determine the appropriate reference frame for the displacement field. This was achieved by identifying the baseline velocity and grouping neighboring images that shared a velocity close to this baseline. The image at the center of the largest segment was then selected as the reference frame. In contrast, in \cite{sala2018musclemotion}, the velocity was used to identify the reference frame by comparing the frame-to-frame calculation values with their adjacent values while simultaneously checking for the lowest absolute value.

In high-throughput drug discovery, where large amounts of data are analyzed, speed and efficiency play a crucial role \cite{mordwinkin2013review}. Hence, we evaluated the efficiency and accuracy of the motion analysis pipeline by down-sampling the images. Our analysis revealed that a complete analysis could be performed in less than a minute on a regular laptop. Furthermore, downsampling the images to 50\% resolution could accelerate the analysis by a factor of three without significant loss of accuracy.

To further evaluate our method, we applied it to a dose escalation study involving Omecamtiv mecarbil and Bay K8644. Our results were consistent with those reported in \cite{gossmann2020integration}, where both compounds were found to increase the peak twitch amplitude.

In this study, we limit the number of algorithms to a few classical algorithms that are available through the OpenCV library. However, more sophisticated methods that exploit deep learning approaches do exist \cite{ilg2017flownet} and could be considered to achieve even better performance. Despite this, the library is fairly efficient and can perform motion analysis within a few seconds on a normal laptop which makes it highly attractive.

One significant advantage of our software compared to other packages is that it is open source, and implemented in Python, Thich allows for a simple installation process and easy integration into existing data processing pipelines.

The typical way you interact with other similar packages is through a graphical user interface. This might be a good way if you want to analyze a few datasets, but it becomes intractable if you need to analyze a large amount of data. Another downside of a purely graphical user interface is that it makes it difficult to integrate into an existing data pipeline where motion analysis would be one component. Furthermore, since most research groups tend to develop their own set of libraries and tools tailored toward the kind of data modalities used in the group, it is often difficult to adopt these tools and integrate them into a new data analysis pipeline.

\section{Conclusion}
We have demonstrated a unified framework for analyzing motion in imaging data of cardiac cells and we have shown how this can be utilized for studying the effect of drugs. 

\section{Acknowledgments}
This work was funded in part by the IDENTIPHY grant from the Research Council of Norway [\#309871/E50].

\clearpage
\bibliographystyle{unsrt}
\bibliography{bibliography.bib}

\appendix

\section{Estimating the reference frame \label{sec:reference_frame}}

\begin{figure}
    \includegraphics[width=\textwidth]{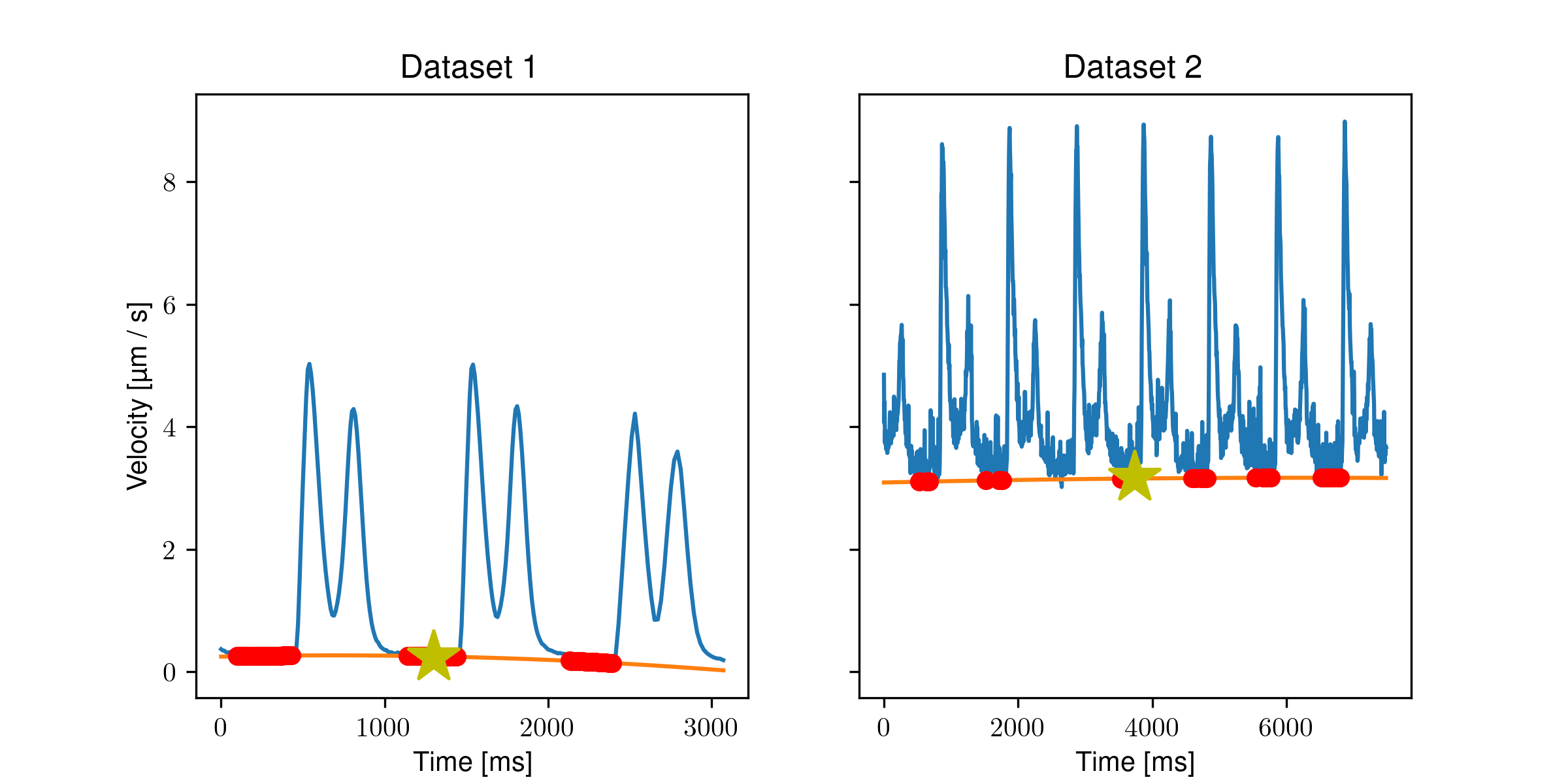}
    \caption{\label{fig:estimate_referece_frame} Details around the estimation of the reference frame for Dataset 1(left) and Dataset 2(right). The blue traces show the estimated velocity traces, and the orange line shows the estimated baseline. The red dots show the points are relatively close to the baseline (within a relative tolerance of 0.005) and the yellow stars shows the selected point for the reference frame.}
\end{figure}

To estimate the displacement, a meaningful reference frame needs to be chosen, preferably before the initiation of contraction and motion such that one can characterize the entire twitch relative to the resting state. However, unless we have extra information, such as when the pacing is triggered, we cannot determine which frame should be chosen as the reference frame without first running the algorithm with an initial guess. The resulting mean trace is highly influenced by the chosen reference though, as the time point with zero displacements will occur at the chosen reference frame.  If this in the middle of contraction, the resulting trace is not clear.   Moreover, finding the correct reference frame from a trace that uses the wrong reference frame might be challenging for the inexperienced. It should, however, be emphasized that there are typically many frames from a image sequence that qualify as good choices of reference frame. The commonality between these is that the cells are at rest with no velocity in this frame. One way to find all the possible reference frames is to compute the velocity for each image relative to the frames around it, and then simply select one of the frames where the norm of the velocity is zero. However, since the overall velocity signal might be noisy, we cannot rely that there will be any time points with zero norm velocity. We can circumvent this by first estimating the baseline velocity and then selecting one of the frames that happen to be on the baseline. To do this in a reproducible way we will group the neighboring frames on the baseline together, find the biggest group and then choose the frame in the center of this group as the reference frame as outlined in Algorithm \ref{fun:estimateRefereceImage}. These groups as well as the chosen reference frame is shown in more details for two selected traces in \cref{fig:estimate_referece_frame}.

\begin{algorithm}
    \caption{estimateReferenceImage($t, v$, reltol=0.005)}\label{fun:estimateRefereceImage}
    \KwIn{$t$ -- Array of time points}
    \KwIn{$v$ -- Array of velocity at each time point.}
    \KwIn{reltol -- Relative tolerance}
    \KwOut{$idx$ -- Index to be used as reference frame}
    Estimate the baseline for the velocity using \cite{mazet2005background}\\
    Find all points where the velocities are close to the baseline (within the given relative tolerance)\\
    Find the biggest connected segment of points that are close to the baseline \\
    \KwRet Point in the center of this segment
\end{algorithm}

\section{Effect of image spacing in the velocity computations \label{sec:spacing}}

\begin{figure}
    \includegraphics[width=\textwidth]{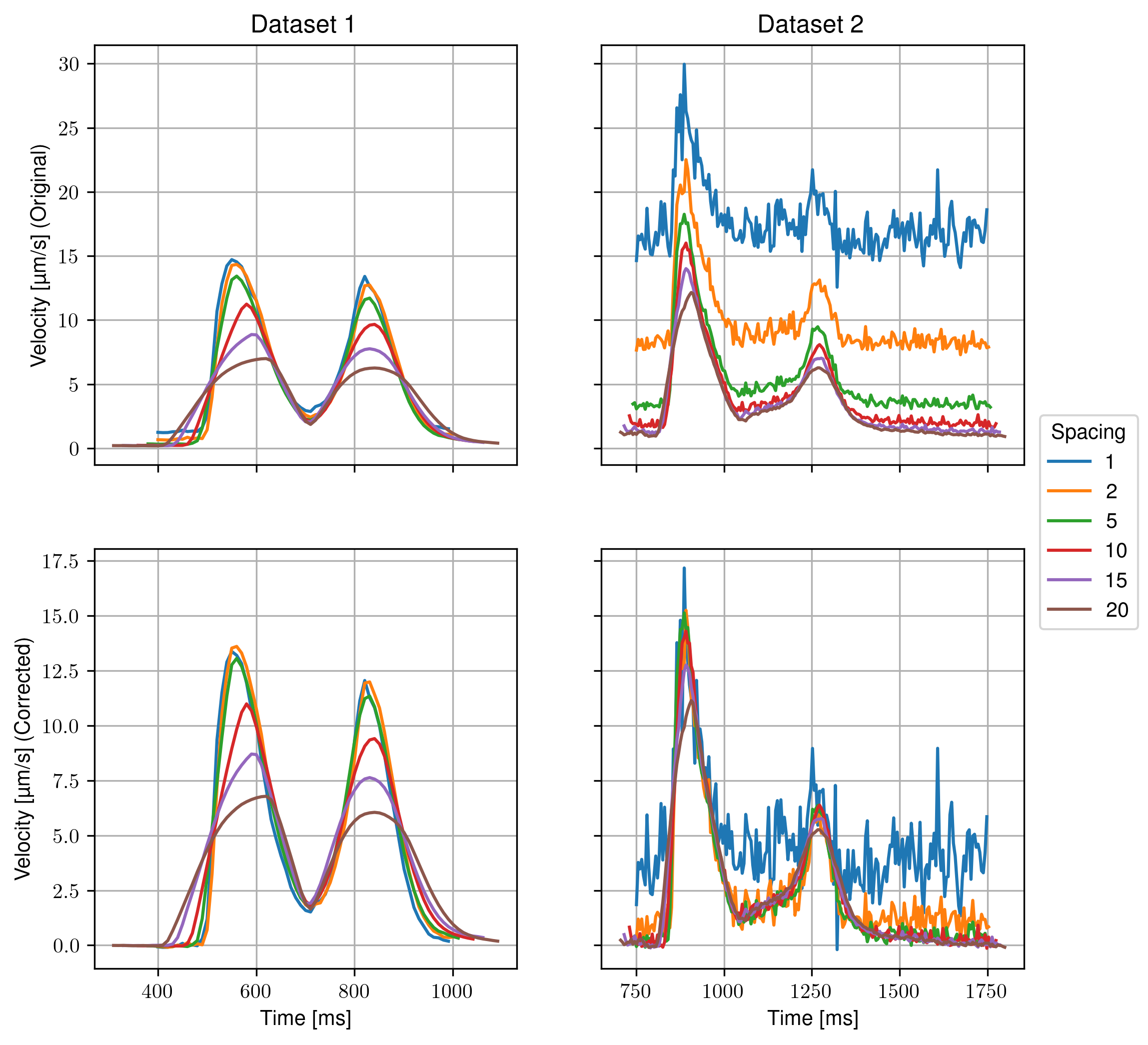}
    \caption{Effect of using different image spacing in the velocity computations\label{fig:velocity_spacing} for Dataset 1 (left) and Dataset 2 (right). Top row shows the original traces when the bottom row shows the traces with baseline correction. Each line represents a different value of the spacing between the frames when computing the velocity. }
\end{figure}

The velocity at a given point $t_k$ is given by the optical flow between two frames in a neighborhood of $t_k$ multiplied by the reciprocal of the time between those two frames, i.e from $t_{k-s}$ to $t_k$ for some $s > 0$, see \cref{eq:velocity_motion}. In this case $s$ is delay between the two frames. We can think about the velocity as the average motion between two successive frames with $s$ frames apart. We will refer to $s$ as the \emph{image spacing}. The image spacing will influence the accuracy as well as the noise level in the velocity traces. \cref{fig:velocity_spacing} shows the first beat of the resulting velocity traces calculated using different amounts of image spacing with and without baseline correction. We notice that even with baseline correction the peak contraction and relaxation velocity is reduced when we increase the spacing. Also, for Dataset 2 we see a significant increase in the level of noise and baseline drifting for smaller image spacing.

In \cref{tab:relative_peak_velocity} we show the peak velocity values relative to the baseline values. We see that the relative peak value declines with an increase in the spacing, however, the value remains more or less the same value when the spacing is between 1 and 5 for Dataset 1, and when the spacing is between 2 and 10 for Dataset 2.

\begin{table}
\centering
\begin{tabular}{lrrrr}
\toprule
Spacing & Dataset 1 & Dataset 2 & $s \cdot dt$ (Dataset 1) & $s \cdot dt$ (Dataset 2) \\

\midrule
1 & 14.0 (0.7/14.7) & 16.6 (14.2/30.8) & 11.9  & 5.0 \\ 
2 & 13.9 (0.4/14.4) & 15.0 (7.9/22.9) & 23.9  & 10.0 \\ 
5 & 13.1 (0.3/13.4) & 14.9 (3.9/18.9) & 59.6  & 25.1 \\ 
10 & 10.8 (0.4/11.2) & 14.2 (2.3/16.5) & 119.3  & 50.1 \\ 
15 & 7.4 (1.5/8.9) & 12.7 (1.7/14.4) & 178.9  & 75.2 \\ 
20 & 3.8 (3.2/7.0) & 10.8 (1.5/12.3) & 238.5  & 100.2 \\ 
\bottomrule
\end{tabular}
\caption{Peak velocity values relative to the baseline velocity for different values of the spacing. Each entry in the table shows the relative peak velocity value for Dataset 1 (column 2) and Dataset 2 (column 3). In parenthesis we show the baseline value followed by the peak value separated by $/$. In column four and five we also display the actual time i milliseconds corresponding to a given spacing for the two datasets.}
\label{tab:relative_peak_velocity}
\end{table}

It is evident that choosing a too low value of the spacing will increase the noise level while choosing a too high spacing will smooth out the signal too much.

When comparing Dataset 1 and 2 we should also take into account that the framerate for the two datasets are different as can be seen in \cref{tab:chip_metadata}. In column four and five in \cref{tab:relative_peak_velocity} we convert the spacing to a physical time between frames based on the frame rate. Choosing a spacing corresponding to a phyiscal time between 25 and 100 milliseconds this seems to give a good trade-off between accuracy and noise accumulation in general and for this case we therefore select a spacing of 5 as the default spacing value for both datasets.

\section{Estimation of motion from fluorescence data}
The displacement and velocity in the results in \cref{fig:full_traces_drug}  are estimated from brightfield images, but we could also estimate this directly from the fluorescence data. There are several benefits to doing so. First of all, we could skip recording the brightfield data which can save time and storage space. Second, we can use the information within the fluorescence data to improve the analysis of the motion data. For example, we can use the fluorescence data to find when the upstroke happens and chop the motion data into beats based on this information.

\begin{figure}
    \centering
   \includegraphics[width=0.45\textwidth]{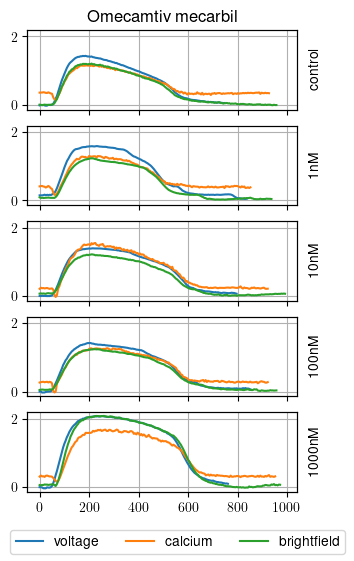}
   \includegraphics[width=0.45\textwidth]{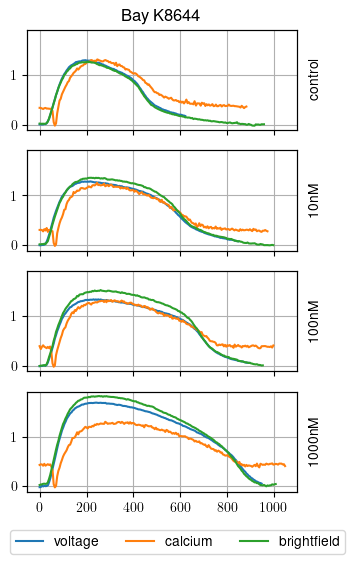}
    \caption{\label{fig:drug_displacement_multimodal}Displacement traces from calcium, voltage and brightfield data}
\end{figure}

We see that the motion estimation also works on the fluorescence data and that it does capture the same beating pattern for all modalities. However, the traces computed from the calcium data are noisier, which might be due to higher pixel values arising from the calcium fluorescence and that the assumption about no change in illumination is not valid.

\end{document}